\documentclass[12pt]{iopart}  

\usepackage{iopams}
\usepackage{epsfig}
\usepackage{amstext}

\usepackage{latexsym}
\usepackage{graphics}

\begin{document}

\paper{Discretisation effects and the influence of walking speed 
in cellular automata models for pedestrian dynamics}

\author{Ansgar Kirchner$^{1,2}$\footnote{Present address:
BERATA GmbH, Gesch\"aftsstelle Hamburg, Nagelsweg 24, 20097 Hamburg}, 
Hubert Kl\"upfel$^{2,3}$,
Katsuhiro Nishinari$^{4,1}$, Andreas Schadschneider$^{1}$,
Michael Schreckenberg$^{2}$}

\address{$^1$ Institut f\"ur Theoretische Physik, Universit\"at zu 
K\"oln, 50923 K\"oln, Germany}

\address{$^2$ Physik von Transport und Verkehr, Universit\"at Duisburg-Essen, 
47048~Duisburg, Germany}

\address{$^3$ TraffGo GmbH, Falkstr. 73-77, 47058 Duisburg, Germany}

\address{$^4$ Department of Applied Mathematics and Informatics,
  Ryukoku University, Shiga, Japan}

\eads{\mailto{aki@thp.uni-koeln.de}, \mailto{kluepfel@traffgo.com}, 
\mailto{knishi@rins.ryukoku.ac.jp}, \mailto{as@thp.uni-koeln.de}, 
\mailto{schreckenberg@uni-duisburg.de}}

\date{\today}%

\begin{abstract}
  We study discretisation effects in cellular automata models for 
  pedestrian dynamics by reducing the cell size. Then a particle
  occupies more than one cell which leads to subtle effects in the
  dynamics, e.g.\ non-local conflict situations.
  Results from computer 
  simulations of the floor field model are compared with empirical findings.
  Furthermore the influence of increasing the maximal walking speed 
  $v_{{\rm max}}$ is investigated by increasing 
  the interaction range beyond nearest neighbour interactions.
  The extension of the model to $v_{{\rm max}}>1$ turns out to be a severe 
  challenge which can be solved in different ways. Four major variants
  are discussed that take into account different dynamical aspects.  
  The variation of $v_{{\rm max}}$ has strong influence on the shape 
  of the flow-density relation. We show that walking 
  speeds $v_{{\rm max}}>1$ lead to results which are in very good 
  agreement with empirical data.
\end{abstract}

\submitto{JSTAT}


\section{Introduction}

The understanding of the dynamical features of pedestrian dynamics has
been the aim of many investigations over the last
years \cite{PedeEvak,dhrev,nagatani}. Recently also approaches
based on  cellular automata (CA) have been suggested.
CA models are discrete in space, time and state variables.
On the one hand, this makes the models ideally suited for large-scale
computer simulations. On the other hand, the discreteness has to
be regarded as an approximation of reality. In this paper we want to
investigate some more fundamental questions related to
general aspects of physical modelling using discrete models.
More specifically, we try to elucidate the qualitative and, 
in some cases, even quantitative influence of the parameters related
to the discretisation. 
As with any other kind of simulation (e.g.\ fluid
dynamics \cite{fluid}), two major decisions have to be made for the
description to be used:
macro- vs.\ microscopic and discrete vs.~continuous \cite{gersh}.
Although we do not consider it explicitly here, performing the 
continuum limit of the discrete models would contribute to a better
understanding of their relation to continuous models.

As we will see, the discreteness leads to some problems that will be
exemplified using the so-called floor field CA model that has been 
introduced in \cite{ourpaper,ourpaper2,KNS}. The big advantage of 
this model compared to other CA approaches for pedestrian dynamics 
\cite{gipps,blue,fukui,min,Kluepfel00} is that the floor field model is able
to reproduce most of the characteristic aspects of pedestrian
dynamics, especially the different collective effects observed
empirically \cite{PedeEvak,dhrev,weid,high}. 
Here, the inclusion of friction effects
allows to describe many important observations, especially
in egress and evacuation scenarios \cite{KNS,KKNSS}. 

There are foremost three basic parameters any CA model is based on:
the interaction range, which in traffic models \cite{css} usually is given
by the maximal velocity $v_{{\rm max}}$, the time scale
$\Delta t$ and the generic length scale $a$, corresponding to the size
of one grid cell. The variation of $\Delta t$ will only cause the
rescaling of all time values such as averaged evacuation times and is
therefore from a theoretical point of view not very interesting. In contrast, 
variations of $v_{{\rm max}}$ and $a$ will turn out to have a strong 
qualitative influence on the {\em dynamical} properties of the model.  

The outline of the paper is as follows: First, we present extensions of 
the basic model (with $v_{{\rm max}}=1$) to larger interaction ranges 
$v_{{\rm max}}>1$.  Four
possible variants of the extension are considered, based on different
aspects of the dynamics, especially the treatment of crossing paths.
Monte Carlo simulations will be presented and compared to empirical
data. Then we investigate the effects of a reduction of the length
scale $a$.  
We will also show that a finer spatial discretisation (by reducing $a$ 
from $a=40$~cm to $a=20$ cm) has strong effects on the dynamics.

But first, we will review the most important characteristics of the
basic model with $v_{{\rm max}}=1$ and $a=40$ cm in the next section.

\section{Cellular automata models}
\label{sec:model}

For pedestrian dynamics, in a natural spatial discretization one
introduces cells of $40\,{\rm cm} \times 40\,{\rm cm}$.
This corresponds to the typical space occupied by a person in
a dense crowd.
Therefore, each cell can then either be empty or occupied 
by (at most) one pedestrian (hard-core exclusion).

Most CA approaches use a parallel (synchronous) update where the
dynamical rules $\mathcal{Z}$ have to be applied to all particles at the 
same time. This introduces a timescale $\Delta t$.
In stochastic CA models the motion to other cells at
each discrete time step $t\to t+\Delta t$ is controlled
by transition probabilities. In the simplest case, which is used
most often, each pedestrian is allowed to move only to one of his/her
four nearest neighbour cells.

The effects of the discretisation and interaction range that we want
to investigate here will appear in any CA model.  However, our testing
ground will be a specific CA, namely the floor field model. It can be
regarded as a two-dimensional generalization of the asymmetric simple
exclusion process (ASEP) (see e.g.\ \cite{gunterrev,evansAlten}) where
the transition probabilities are determined {\em dynamically} through
a coupling to so-called {\em floor fields} \cite{ourpaper,ourpaper2}. In
turn, these floor fields are changed by the motion of the pedestrians
which leads to feedback effects.  This way of implementing the interactions
is inspired by the form of communication in insects societies (e.g.\ ants)
and the interaction can be described as {\em virtual chemotaxis}. 
Generically a movement in the direction of larger fields is preferred.
The key point is that an interaction that is long ranged in space
can be translated into a local interaction, but with ``memory''.
Note that in contrast to chemotaxis and trail formation
\cite{benjacob1,ants,trail1,trail2} in our model the trace is only virtual, 
i.e.\ can not be observed empirically.
We do not give a full definition and
discussion of the floor field model, which can be found in
\cite{ourpaper,ourpaper2,KNS,aki,kluepfelthesis}, but focus on the 
aspects that are important in the following.

The floor field consists of two parts, a static and a dynamic field.
Both are discrete and we can imagine the field strength as the
number of virtual pheromones present at a site.
The fixed {\em static floor field} $S$ is not changed by the presence
of the pedestrians. $S$ is used to specify regions of space which are
more attractive, e.g. exits or shop windows. In case of
an evacuation simulation the static floor field $S$
describes the shortest distance to the closest exit door measured in steps,
i.e.\ the number of lattice sites to the exit \cite{ourpaper2}. 
Steps can be carried out to the four next neighbour sites.
The field values increase  towards the door. 

Collective effects like lane formation or herding 
\cite{PedeEvak,dhrev,social,panic}
that are based on a long
range interaction between pedestrians can be taken into account by a
dynamically varying floor field $D$ \cite{ourpaper}.
The {\em dynamic floor field} $D$ is a virtual trace left by the
pedestrians and has its own dynamics, i.e.\ diffusion and decay.  It
is used to model different forms of interaction between the
pedestrians. At $t=0$ the dynamic field is set to
zero for all sites $(x,y)$ of the lattice, i.e.\ $D_{xy}=0$. Whenever
a particle jumps from site $(x,y)$ to one of the four neighbouring
cells, $D$ at the origin cell is increased by one: $D_{xy}\to D_{xy}+1$. 
Furthermore, the dynamic floor field is time dependent, it has diffusion 
and decay controlled by two parameters $0\leq \alpha,\delta\leq 1$,
which leads to broadening and dilution of the trace \cite{ourpaper}.

The coupling of the particles to the static and dynamic floor field is 
controlled by coupling constants $k_S$ and $k_D$, respectively.
A strong coupling to the static field implies an almost deterministic
motion in the direction of larger fields. On the other hand, if
the coupling to the dynamic field dominates, herding effects become
important where pedestrians blindly follow others as it may happen
in the case of panic.

An important extension that makes the dynamics more realistic
is the concept of {\em friction} \cite{KNS,KKNSS}. It has been
implemented in other models in a direct way as contact friction
\cite{panic}. However, in our approach it concerns the behavior of 
two (or more) pedestrians that try to move to the same cell.
With some probability $\mu$, called friction parameter,
none of the particles is allowed to move, whereas with probability
$1-\mu$ one of the particles is chosen randomly to occupy the
target cell. This leads to effects similar to arching in 
granular materials \cite{panic,arching}.


\section{Extensions to $v_{{\rm max}}>1$}
\label{vervmax}

\begin{figure}[h]
\begin{center}
\includegraphics[width=.6\textwidth]{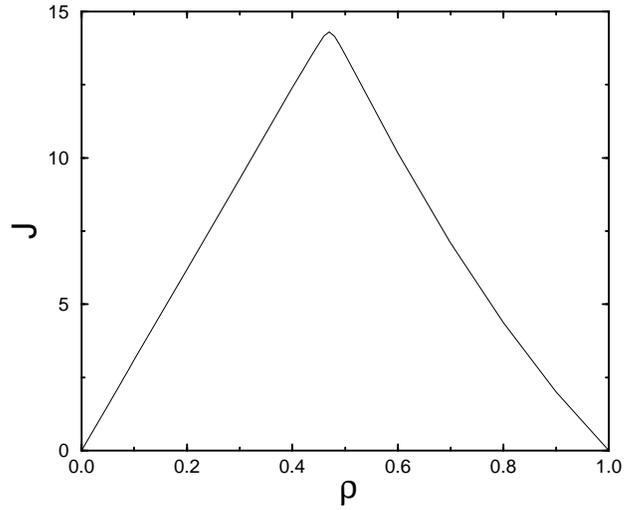}
\end{center}
\caption[]{Fundamental diagram (flow $J$, measured in units of
$1 / \Delta t$, vs.\ density $\rho$, measured in units of 
$\rho_{\rm max}=1/a^2=6.25$ m$^{-2}$))
  for a corridor of $93\times 33$ lattice cells of size $a^2=40\times 40$
  cm$^2$ with $v_{{\rm max}}=1$ 
  in the so-called ordered regime ($k_S=10.0$ and $k_D=0.0$).  
  The static floor field $S$ is calculated using a Manhattan metric. 
  In the direction of the flow periodic boundary conditions are used.
  For details see \cite{ourpaper2,aki}.}
\label{classfund}
\end{figure}

In the definition of the model given above, particles are only allowed
to move to unoccupied {\em nearest} neighbour sites in one time step
$t\to t+\Delta t$, i.e.\ $v_{{\rm max}}=1$~cell/time step. This constraint
corresponds to a maximum empirical walking speed of $v_{{\rm emp}}=1.3$ m/s
\cite{weid,high}, if the time scale $\Delta t$ is identified with
$0.3$ s. The choice $v_{{\rm max}}=1$ has the advantage of simplicity
and high simulation speed.  It also reproduces most experimentally
observable phenomena.  Nevertheless, there are reasons for the
introduction of higher walking speeds $v_{{\rm max}}>1$ in some cases:
\begin{enumerate}
\item Fig.~\ref{classfund} shows the fundamental diagram of the flow
  along a corridor in the $v_{{\rm max}}=1$ case of large coupling 
  $k_S$ and $k_D=0$. \footnote{The empirical
  unit of the specific flow is pedestrians$/$(meter$\cdot$second), the
  natural unit of the model: particles$/$($a\cdot\Delta t$) 
  (with cell length $a=$40 cm and time scale $\Delta t$).}   
  It is nearly
  symmetric with maximal flow at $\rho\approx\frac{1}{2}$.
  However, experimental data point to a non-symmetric fundamental
  diagram with maximal values of the flow for densities
  $\rho<\frac{1}{2}$ \cite{weid,high}.
\item The interaction horizon is not isotropic since pedestrians react
  mainly to stimuli in front of them. This anisotropy can 
  be better taken into account in a model with larger interaction range 
  where it is generated dynamically through the exclusion principle.
\item Although the velocity distribution of a pedestrian is sharply
  peaked around 1.3~m/s, higher walking speeds are frequently observed
  \cite{weid,high}. 
\item In a model with $v_{{\rm max}}>1$, a realistic distribution of
  different walking speeds of the particle ensemble can be implemented
  in a simple way.
\end{enumerate}
Certainly the most relevant aspects are the first two. The other two 
requirements could be taken into account in the case $v_{{\rm max}}=1$ 
by introducing individual randomization probabilities, determined 
by the coupling constants $k_S$ and $k_D$, or a rescaling of the 
relevant time scale.
In the following extensions of the model to $v_{{\rm max}}>1$ will
be presented. The possible variants of the update algorithm and their
dynamical properties will be discussed in detail and compared to
experimental data.

 
\subsection{Classification of the model variants}

In the case of $v_{{\rm max}}=1$, the particles are allowed to cover a
maximal distance of $a=40$ cm (corresponding to a movement to next
neighbouring sites) per time step.  Then for higher walking speeds
$v_{{\rm max}}\geq 2$ the particles should be able to cover a distance 
of $v_{{\rm max}}\cdot a$ in one time step. 
This does not necessarily correspond to higher velocities, e.g.\
when combined with a reduction of the cell size or change of time scale,
which is the case mainly considered here.
Then, due to the two-dimensional nature of the motion, the choosen path 
is not necessarily straight. Also it has to be taken into account that 
now the paths of different pedestrians may cross. This requires a 
generalization of the concept of friction. In the following we develop a 
formalism which allows to describe the different possible generalizations.

\begin{figure}[h]
\begin{center}
  \includegraphics[width=.3\textwidth]{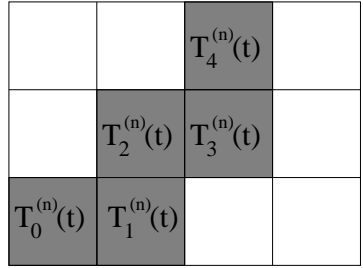}
\end{center}
\caption[]{Graphical representation of the desired trajectory
$T^{(n)}_{l}(t)$ of pedestrian $n$ for the case $v_{{\text max}}\geq 4$.}
\label{fig-tmat}
\end{figure}

At the beginning of a time step $t$ the position of each
particle $n\in\{1,\ldots,N\}$ is denoted by $T^{(n)}_{0}(t)$. Each
particle then chooses, corresponding to its desired movement, a
trajectory $T^{(n)}(t)=\{T^{(n)}_{0}(t),\ldots,T^{(n)}_{v_{{\rm max}}}(t)\}$
consisting of $v_{{\rm max}}+1$ lattice sites.  The single elements of
$T^{(n)}(t)$ are determined by applying the update rules $\mathcal{Z}$
for the $v_{{\rm max}}=1$ case successively $v_{{\rm max}}$ times.  Here,
every arbitrary element $T^{(n)}_{l}(t)$ ($l>0$) corresponds to a target
site $\mathcal{Z}$ of the particle $n$ that has advanced virtually 
to site $T^{(n)}_{l-1}(t)$ (see Fig.~\ref{fig-tmat}):
\begin{equation}
T^{(n)}_{l}(t)=\mathcal{Z}(T^{(n)}_{l-1}(t)).
\end{equation}
Therefore, $T^{(n)}_{l}(t)$ can only be one of the four next neighbour
sites of $T^{(n)}_{l-1}(t)$ or $T^{(n)}_{l}(t)=T^{(n)}_{l-1}(t)$.  
Thus, every particle accesses iteratively at
most $v_{{\text max}}$ target sites. It is
important to notice that the $v_{{\text max}}$ target sites
and thus the trajectories are determined using the 
configuration of the system at 
time $t$ through $\{T^{(1)}_{0}(t),\ldots,T^{(N)}_{0}(t)\}$.

All desired trajectories can be summarized in the matrix
\begin{equation}
T(t)=
\left(
\begin{array}{c c c}
T^{(1)}_{0}(t) & {\cdots} & T^{(1)}_{v_{{\rm max}}}(t)\\
\vdots & \ddots & \vdots \\
T^{(N)}_{0}(t) & \cdots & T^{(N)}_{v_{{\rm max}}}(t)
\end{array}
\right)
\;,
\end{equation} 
which has to be stored at every time step of the simulation. Two basic 
assumptions of the model are reflected by the matrix elements of $T(t)$:
\begin{itemize}
\item Hard-core exclusion of particles:
$T^{(i)}_{0}(t) \ne T^{(j)}_{0}(t)$ for all $i \ne j$, i.e.\
every lattice site can be occupied by at most one particle.
\item Parallel updating: 
$T^{(i)}_{v_{{\rm max}}}(t)\ne T^{(j)}_{0}(t)$ for all $i \ne j$.
This expresses the reaction time of the particles.
No particle can occupy a lattice site at time step $t+1$, which was occupied 
at time step $t$ by another one.
\end{itemize}
These properties are special cases of the more general rule
$T^{(i)}_{n}(t) \ne T^{(j)}_{0}(t)$ for all $i \ne j$ and $n=0,1,
\ldots,v_{{\rm max}}$.
They restrict the allowed values of the first and last column of
$T(t)$, i.e.\ the allowed positions at time $t$ and $t+\Delta t$.
The remaining $v_{{\rm max}}-1$ columns correspond to intermediate
states that are not explicitly realized as configurations. 
It might happen that two or more desired trajectories cross or that
even one of the intermediate states violates the hard-core exclusion
principle. In the following we will denote such situations as 
{\em conflicts}. In principle one could ignore  the existence of
conflicts since they correspond only to virtual intermediate states.
However, in order to make the dynamics more realistic it is necessary
to take them seriously. Then a method for the resolution of conflicts
is required. In the following we will discuss four major variants.

\subsubsection{Hop or stop}
\label{sub_hop}

The simplest variant ignores the intermediate positions 
$T^{(n)}_{l}(t)$ ($l=1,\ldots,v_{{\rm max}}-1$) and takes
only conflicts in the desired new positions $T^{(n)}_{v_{{\rm max}}}(t)$
into account. If $k$ particles ($k\geq 2$) have chosen the same
target cell, i.e.
\begin{equation}
T^{(i)}_{v_{{\rm max}}}(t)=T^{(j)}_{v_{{\rm max}}}(t),\quad{\rm for\ all}\; 
i,j\in\{1,\ldots,k\}\;,
\end{equation}
one particle $i$ is chosen by a probabilistic method (i.e.\ in
random order) and is allowed to move to target site 
$T^{(i)}_{v_{{\rm max}}}(t)=:T^{(i)}_{0}(t+1)$. 
The other $k-1$ particles $j\neq i$ involved in the conflict are not
allowed to move and have to remain at their origin sites, i.e. 
$T^{(j)}_{0}(t+1)=T^{(j)}_{0}(t)$. Practically, this can be easily
implemented by allowing all particles to move sequentially by choosing
randomly a permutation $\sigma$ of the particle indices $j$ that
determines the order $\sigma(j)$ of updating\footnote{This implies that
the particles move in the order $\sigma^{-1}(1),\sigma^{-1}(2),\ldots, 
\sigma^{-1}(N)$.}. 

This is the simplest form of conflict resolution between particles. The 
overall movement takes place without consideration  of the trajectories 
$T^{(n)}(t)$. Crossing trajectories and jumping over already 
advanced particles are permitted in order to keep the spirit of
the parallel updating scheme.

\subsubsection{Move as far as possible} 

This variant of conflict resolution is similar to the previous one. 
Again the particles are allowed to move in random order described
by a randomly chosen permutation $\sigma$.
However, now other particles that have already reached their target sites 
cannot be jumped over by following particles anymore.
A particle $n$ is able to proceed along its trajectory $T^{(n)}(t)$, 
as long as $T^{(n)}_{l}(t)$ ($l\in\{1,\ldots,v_{{\rm max}}\}$) is not 
already occupied by another particle.

In this variant crossing trajectories are still possible, but jumps over
particles that have already reached their target sites are not allowed.

\subsubsection{$v_{{\text max}}$ sub-steps}
\label{sec_substeps}

In this variant of the update the particle movement along the
trajectories $T_n(t)$ is subdivided into $v_{{\rm max}}$ sub-time
steps where moves to nearest-neighbour cells are allowed. 
If in one sub-time step $l$, two or more (at most four)
particles try to move to the same cell, i.e.
\begin{equation}
T^{(i)}_{l}=T^{(j)}_{l},\quad {\rm for\ any}\; i,j,
\end{equation} 
only one randomly chosen particle is allowed to move forward to this
site. All other particles then try to reach the site during the next
sub-time step $l+1$.

In all sub-time steps of this update version conflicts between
particles are taken into account and solved. Therefore, it is the
easiest way to incorporate the concept of friction
\cite{KNS} into the model with $v_{{\rm max}}>1$. Otherwise, this
method leads to increased simulation times due to the successive
treatment of the sub-time steps. Note, that the dynamics of
this variant is different from the genuine $v_{{\rm max}}=1$
-case, since the trajectories $T^{(n)}$ are fixed during the whole time step.

\subsubsection{No crossing paths}

In this last version, no particle is allowed to cross the trajectory of
a particle that has already moved. The particles move in random
order. If the path of particle $n$ crosses that of a particle $n'$
that has already moved (i.e.\ $\sigma(n')<\sigma(n)$),
\begin{equation}
T^{(n)}_{l}=T^{(n')}_{k}\;,\;{\rm with}\;\;l,k\;{\rm arbitrary},
\end{equation}
then 
\begin{equation}
T^{(n)}_{0}(t+1)=T^{(n)}_{l-1}(t)\;.
\end{equation}
$k$ is in this case assumed to be the smallest index in the trajectory
of $n'$ for which this condition holds. Therefore, in this variant
a particle is allowed to move until it reaches a cell that is 
either occupied or has been passed by previous particle that has
already been moved at the same time step.

\subsubsection{General aspects of the variants}

In all variants, at the end of each time step the field values of the 
dynamic field $D$ at all lattice sites of the
trajectories traversed by particles are increased by one. Afterwards
$D$ is modified by its decay and diffusion dynamics as described in
\cite{ourpaper,ourpaper2}. 

\begin{figure}[h]
\begin{center}
  \includegraphics[width=.55\textwidth]{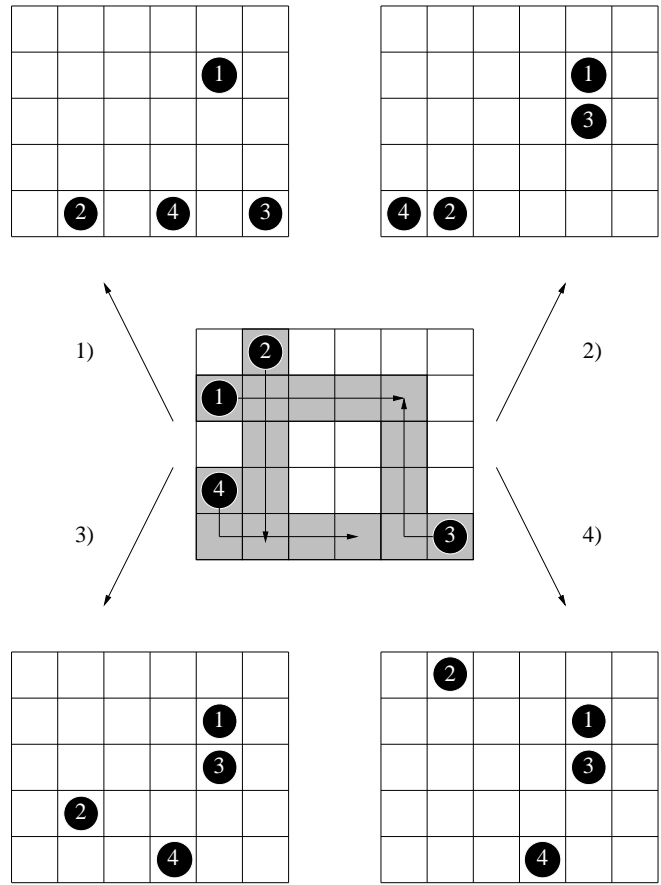}
\end{center}
\caption[]{Graphical representation of all variants of conflict resolution 
  in the case $v_{{\rm max}}=4$. The trajectories $T^{(n)}(t)$ of the
  particles are shaded gray. The numbers of the particles correspond
  to the random order (given by the permutation $\sigma$) in which the
  particles execute their steps. In variants 1), 2) and 4)
  this order is designated once. In the variant 3) one has to choose
  a new random order in every single sub-time step. 
}
\label{speed_1}
\end{figure}

Each of the four update versions is stressing different aspects
of the two-dimensional motion on the lattice. Fig.~\ref{speed_1}
shows a representation of all different methods for conflict
resolution.
All four variants reduce to the basic model described in Sec.~\ref{sec:model} 
in the case of $v_{{\rm max}}=1$.

The characteristic time-scale $\Delta t$ should not be influenced by the 
choice $v_{{\rm max}}>1$. Therefore, for a given value of $v_{{\rm max}}$ 
the corresponding maximum empirical walking speed can be calculated via
\begin{equation}
v_{{\rm max}}^{{\rm (emp)}}=v_{{\rm max}}\cdot\frac{a}{\Delta t}
=v_{{\rm max}}\cdot\,1.3\, {\rm m}/{\rm s}
\label{speed_max}
\end{equation}
with $a=40$ cm and $\Delta t=0.3$ sec.
 
In the following, the consequences of higher walking speeds on particle flows
will be investigated and the differences between the four variants
are discussed. Afterwards, the dependence of the evacuation time for a room on
$v_{{\rm max}}$ is investigated.


\subsection{Simulations}

The particle flows $J=\rho \langle v_x \rangle$ presented in this
section are measured in a corridor of size $X \times Y=93\times33$
with periodic boundary conditions in $x$-direction. The motion
in $y$-direction is limited by walls. The static floor field $S$ increases 
from left to right, i.e.\ $S_{x,y}=x$ \cite{KKNSS}. We remind the
reader, that in the floor field model a motion in the direction
of increasing field strength is preferred. The lattice was initialized 
by randomly distributing $N$ pedestrians corresponding to a density
$\rho=N/(X\cdot Y)$. 
If not stated otherwise, the coupling $k_D$ to the dynamical field
vanishes ($k_D=0$) and therefore the dynamical parameters of $D$ can
be chosen as $\alpha=0$ and $\delta=1$.

\subsubsection{Fundamental diagrams for different values of $v_{{\rm max}}$}

\begin{figure}[h]
a)\hspace{0.485\textwidth} b)\vspace{-0.7cm}
\begin{center}
\includegraphics[width=0.495\textwidth]{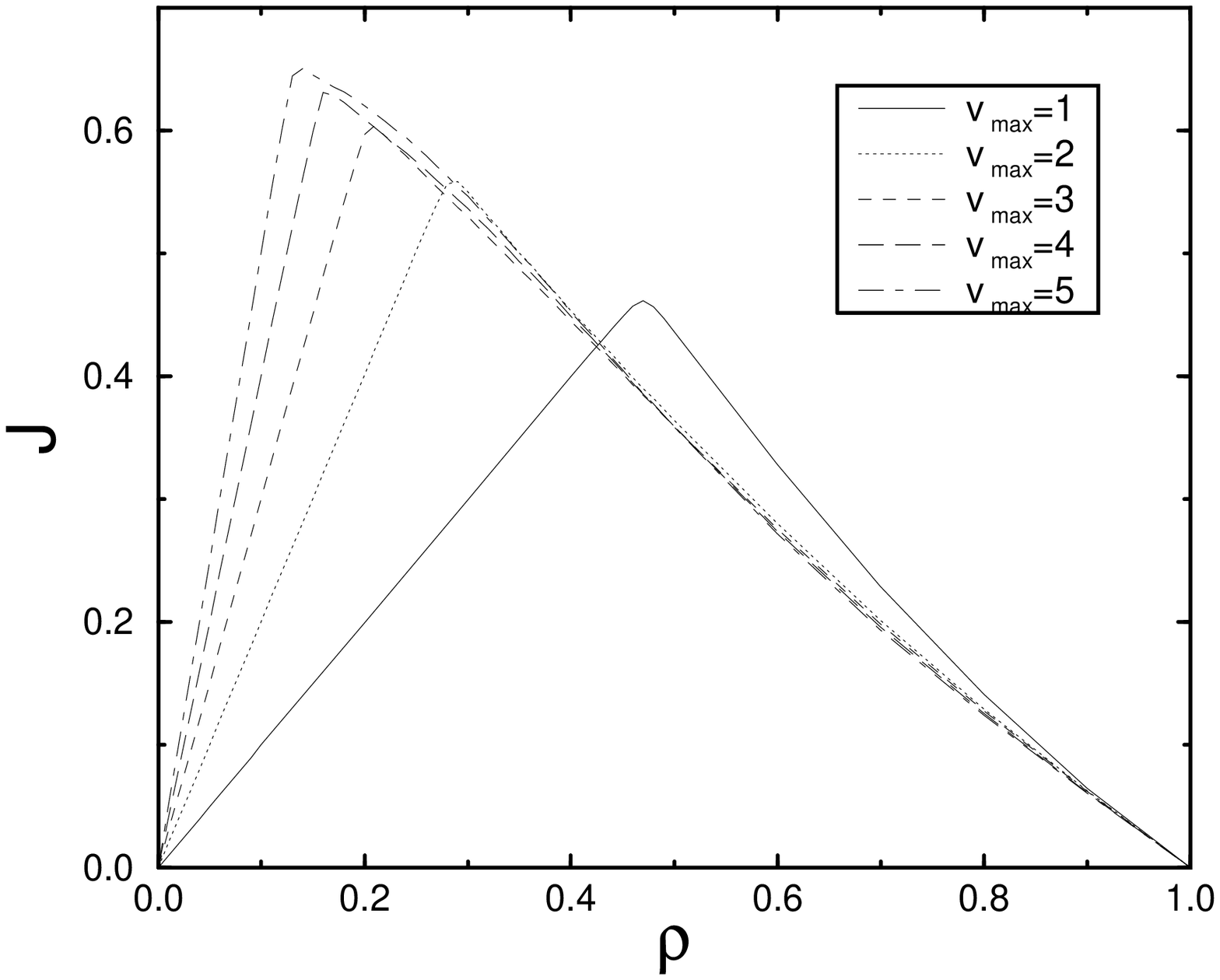}
\hfill
\includegraphics[width=0.495\textwidth]{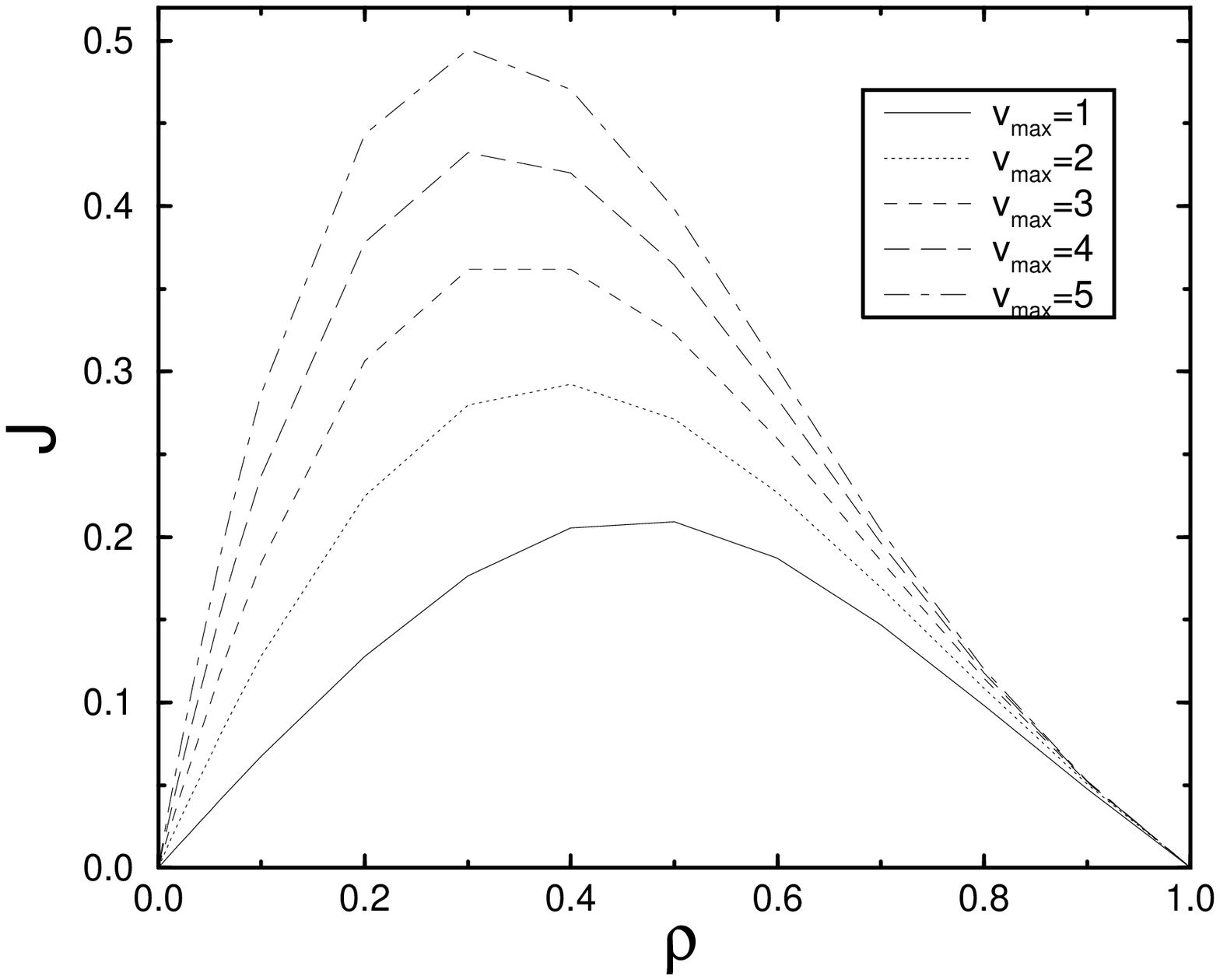}
\end{center}
\caption{ Fundamental diagrams for $v_{{\rm max}}=1,\ldots,5$.
{\bf a)} Variant 4 ({\it no crossing paths}) with
  $k_S=10.0$ and $k_D=0.0$; {\bf b)} Variant 2 ({\it move as far as
    possible}) with $k_S=2.0$ and $k_D=0.0$.}
\label{v_fig_1}
\end{figure}

Fig.~\ref{v_fig_1} shows fundamental diagrams  for $v_{{\rm max}}=1,\ldots,
5$ for two of the variants discussed above. 
In Fig.~\ref{v_fig_1}a the flow values $J$ are shown for the
case of a dominating coupling to the static field
($k_S\gg k_D$, corresponding to the ordered regime \cite{ourpaper2}).
The update is based on the {\it no crossing paths} variant. The
maximal flow is reached at smaller particle
densities for growing $v_{{\rm max}}$. The maximal flow is
also increasing with $v_{{\rm max}}$. An unexpected property of this 
variant is that for densities $\rho>0.4$ the flow $J$ is largest in the
case $v_{{\rm max}}=1$. This can be
explained by the blockage of wide parts of the lattice, resulting from
the non-traversable trajectories of the particles. The influence of
this blockage increases with $v_{{\rm max}}$ due to an increased 
`effective density'. The faster a
pedestrian moves, the more cells she or he blocks. In this respect,
the {\em no crossing paths} scheme leads to the highest effective
densities (Fig.~\ref{v_fig_1}).

Because of the strong coupling to the static field $S$, which suppresses 
swaying in $y$-direction, the fundamental diagrams look very similar
to those of the ASEP or other traffic models in the deterministic
limit (Fig.~\ref{v_fig_1}a).
If one chooses, like in
Fig.~\ref{v_fig_1}b for the second version of the update, a smaller
coupling strength ($k_S=2$), the effective walking speed $v_{{\rm eff}}$ 
of the particles is decreasing.  
For $k_s\to \infty$ the effective walking speed of a single pedestrian 
is $v_{\rm eff}\to v_{{\rm max}}$, corresponding to a
deterministic motion. On the other hand, for $k_s=0$ the particle
performs an isotropic random walk since the information about the
preferred direction through the static floor field is not taken
into accout.
Therefore, for $k_S=2.0$ the shape of the fundamental diagrams is becoming
smoother (Fig.~\ref{v_fig_1}b). The maximal values of $J$ are
smaller with a slight shift of the maximum to higher densities
(for $v_{{\rm max}} >1$) and the slopes in
the regions of small and high densities decrease.

\begin{figure}[h]
a)\hspace{0.485\textwidth} b)\vspace{-0.7cm}
\begin{center}
\includegraphics[width=0.495\textwidth]{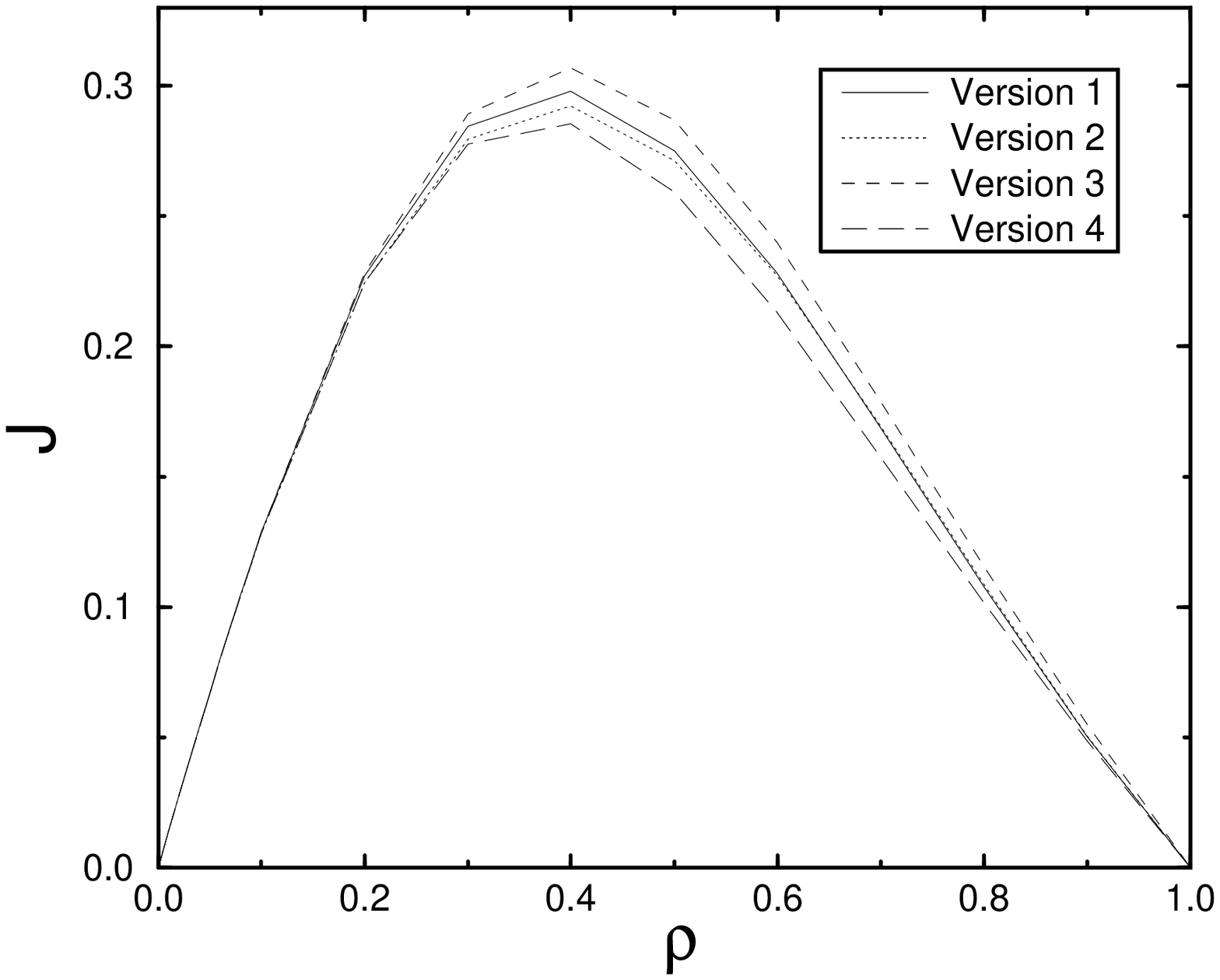}
\hfill
\includegraphics[width=0.495\textwidth]{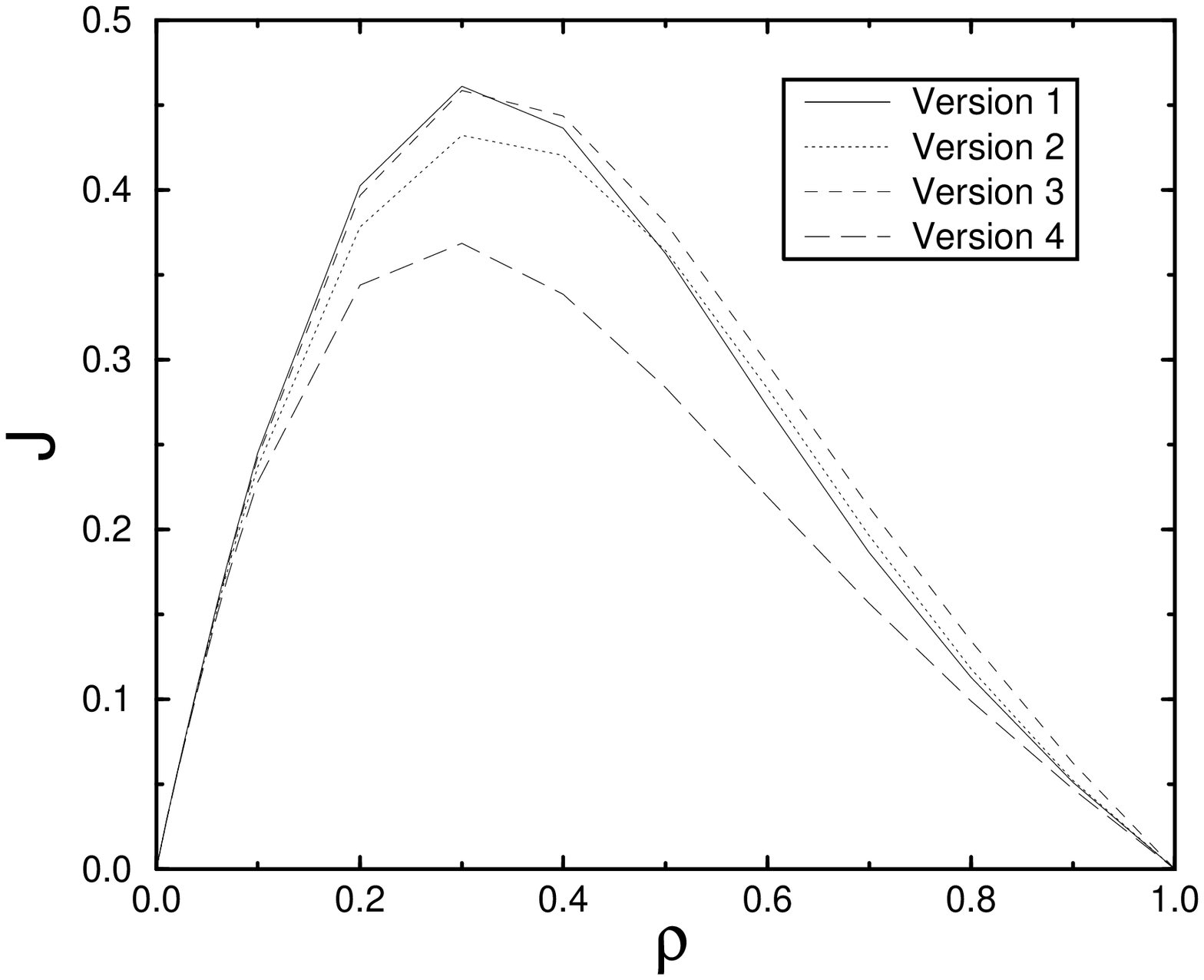}
\end{center}
\caption{Fundamental diagrams for all four update variants with $k_S=2.0$ 
and $k_D=0.0$: {\bf a)} $v_{{\rm max}}=2$; {\bf b)} $v_{{\rm max}}=4$.}
\label{v_fig_2}
\end{figure}

In the following, the consequences of the different conflict
resolution schemes on the flow are discussed.  Fig.~\ref{v_fig_2} 
shows the fundamental diagrams of all four variants for
the velocities $v_{{\rm max}}=2$ and $v_{{\rm max}}=4$ and $k_S=2.0$.
For $v_{{\rm max}}=2$ the differences between the variants are not
strongly pronounced (Fig.~\ref{v_fig_2}a). But the flow values $J$
are highest for the {\it $v_{{\rm max}}$ sub-steps } and smallest for
the {\it no crossing paths} scheme in all density regions.

However, for $v_{{\rm max}}=4$ the differences between the variants
are easy to spot. For particle densities $\rho>0.1$ the {\em no
  crossing paths} scheme produces much smaller flow values than all
other variants, because of the non-traversable trajectories which are
four sites long. The third version of the update ({\em sub-steps}) 
leads to the highest flows in all density regions since the resolution of
conflicts in sub-time steps enables some particles to move forward in
a later sub-time step.

\begin{figure}[h]
\begin{center}
\includegraphics[width=0.55\textwidth]{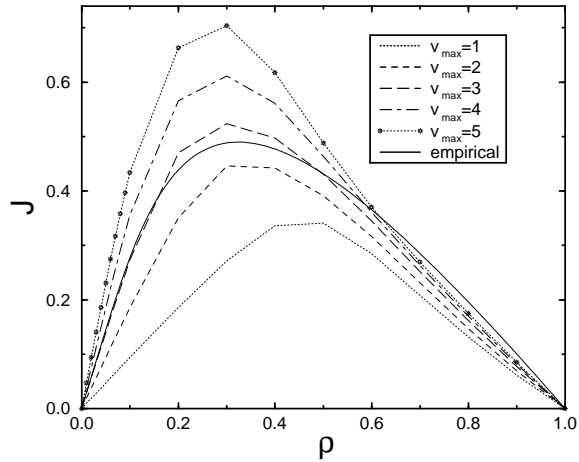}
\end{center}
\caption{Comparison of fundamental diagrams of the third update version 
and the velocities $v_{{\rm max}}=1$ to $v_{{\rm max}}=5$ with experimental 
data from \cite{weid} ($k_S=4.0$ and $k_D=0.0$).}
\label{v_fig_3}
\end{figure}

\begin{figure}[h]
\begin{center}
\includegraphics[width=.65\textwidth]{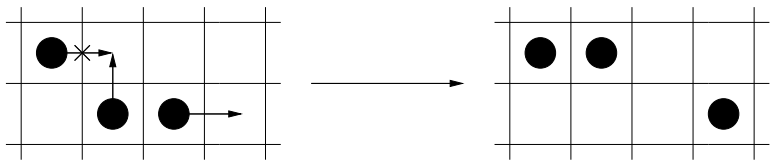}
\end{center}
\caption[]{Decrease of the flow due to lane changes.}
\label{v_fig_4}
\end{figure}

Finally, we compare the simulation results with empirical data.
Fig.~\ref{v_fig_3} shows a comparison of fundamental diagrams of the
{\it $v_{{\rm max}}$ sub-steps } variant (see Sec.~\ref{sec_substeps})
with a fit of empirical flow-density relations \cite{weid}.  
The data for the specific flow
has been transformed to the units of the model.
The fundamental diagram for $v_{{\rm max}}=3$ is in good
agreement with the experimental data. In particular, the density
values for the highest flow match each other very well. Since
$k_S=4.0$, the particles have a reduced effective walking speed
$v_{{\rm eff}} < v_{{\rm max}}$. In the model, this results in a 
diversification of
the flows for particle velocities between $v_{{\rm max}}=3$ and lower
values. For high density regions $\rho>0.6$ the model yields
flow values which are too small. The reason are lane
changes of particles especially at higher densities. If the site in front 
of a particle is occupied, lane changes are
possible. Such a changing does not increase the flow in positive
$x$-direction, but often inhibits the movement of another particle
(see Fig.~\ref{v_fig_4}).
The fundamental diagrams of the other $v_{{\rm max}}$ values do not
agree as well with the empirical data.

\subsection{Simulation of the evacuation of a room}

As a second scenario which helps to understand the effects
of $v_{{\rm max}}>1$ we investigate its influence on evacuation times. 
The room to be evacuated is a square of $63\times63$ 
cells. The exit has a width of one cell and is located at the center 
of one wall. Initially the particles are distributed randomly and then
try to leave the room due to the information they get through the floor 
fields. This scenario has been studied in detail in \cite{ourpaper2} where
three different regimes ({\em ordered} for $k_S \gg k_D$, {\em disordered} 
for $k_D \gg k_S$, and {\em cooperative}) have been identified.

\begin{figure}[h]
a)\hspace{0.482\textwidth} b)\vspace{-0.6cm}
\begin{center}
\includegraphics[width=0.495\textwidth]{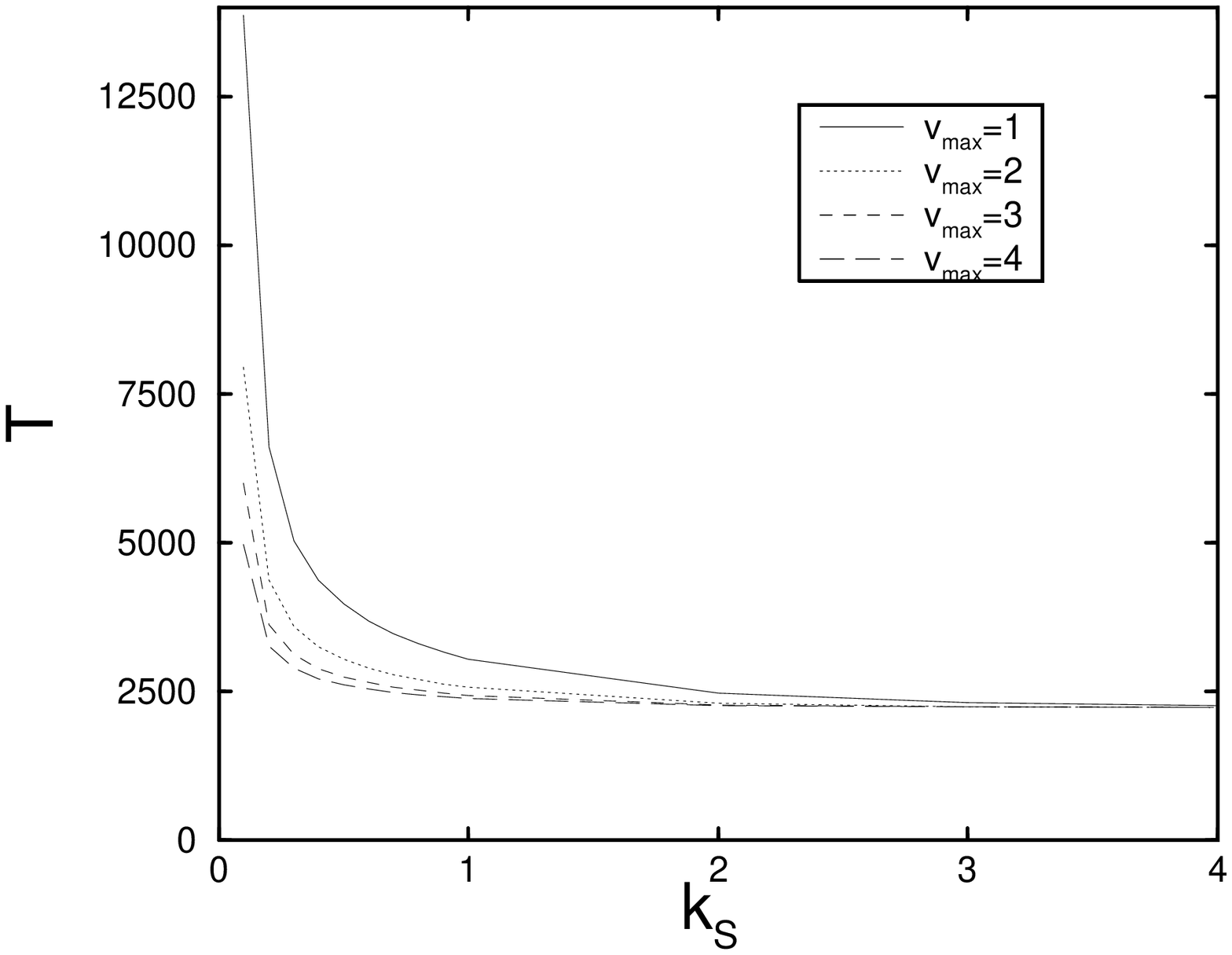}
\hfill
\includegraphics[width=0.495\textwidth]{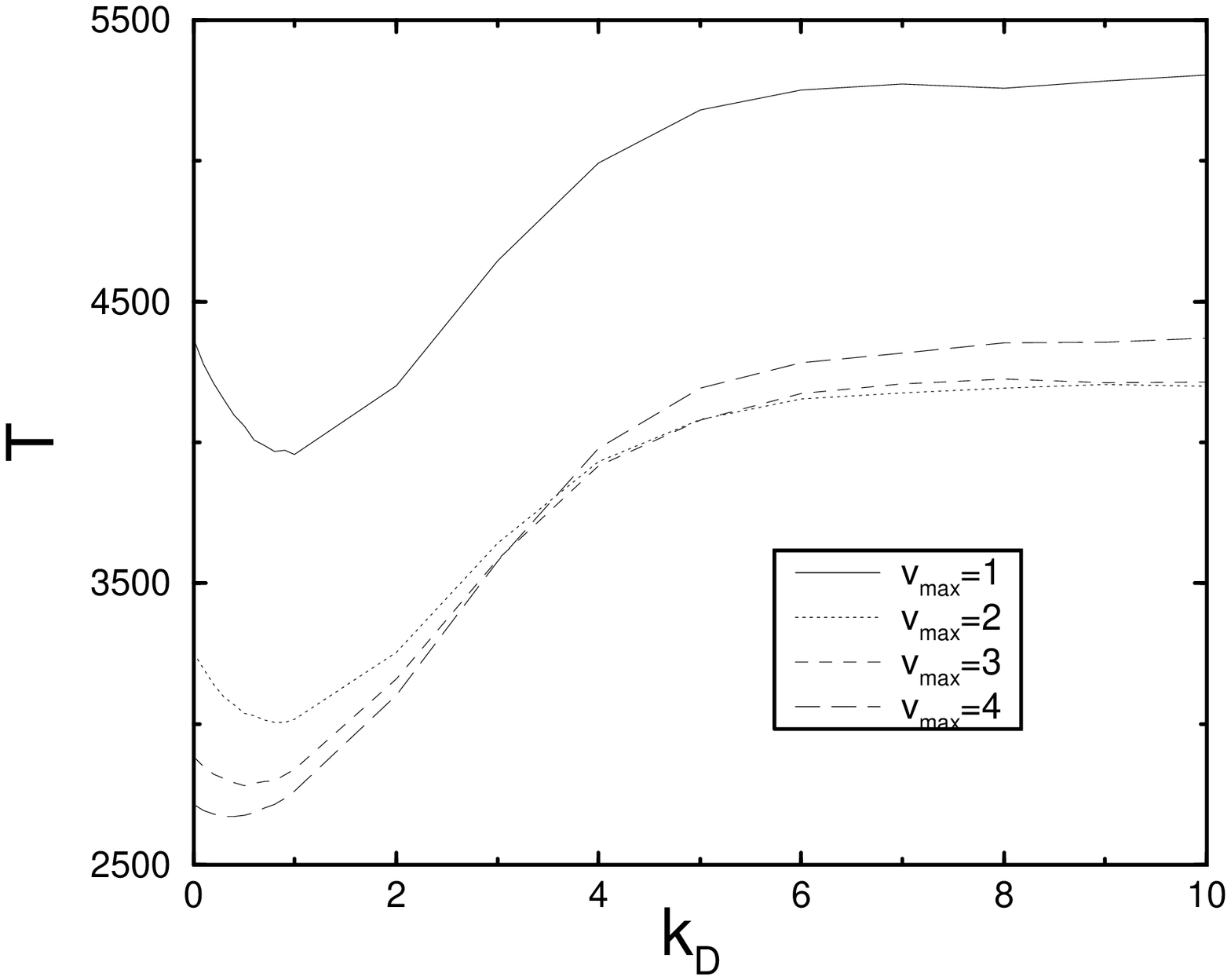}
\end{center}
\caption{Averaged evacuation times for a room of size $63\times 63$ sites 
and an exit of one cell. The particle density at the beginning of the 
evacuation is $\rho=0.3$ (1116 randomly distributed particles). Different 
$v_{{\rm max}}$-values in the second version of the update ({\it move as 
far as possible}).  {\bf a)} $k_D=0.0$; {\bf b)} $k_S=0.4$ and 
$\delta=0.3$ and $\alpha=0.0$ as parameters for the dynamics of $D$.}
\label{v_fig_5}
\end{figure}

Fig.~\ref{v_fig_5} shows averaged evacuation times for different combinations 
of the coupling parameters $k_S$ and $k_D$ for the {\it move as far as
  possible} variant of the update. All other schemes  show the same
qualitative behaviour. The particle velocities are $v_{{\rm max}}=1,
\ldots,4$.
Fig.~\ref{v_fig_5}a can be interpreted in the following way: If the
coupling strength to the static field $S$ is very small ($k_S\ll 1$), 
the particle motion resembles a random walk. But even if
there is only slightly directed movement, the variance in lateral
direction is increased with $v_{{\rm max}}$. Therefore, the particles
can reach the exit by chance faster for increased $v_{{\rm max}}$,
resulting in decreased evacuation times.  In the limit $k_S\to \infty$
the particles move on the shortest possible way to the exit, the
movement becomes deterministic. For higher particle densities just
after the evacuation starts a big jam forms in front of the exit.
Therefore, the averaged evacuation times are in this case determined
by the particle number (via the flow through the door) rather than the
maximum walking speed. Thus, for higher densities and $k_S\to \infty$
the evacuation times using the {\em move as far as possible} strategy
(see Fig.~\ref{v_fig_5} top) are the same for all $v_{{\rm max}}$ values.

A coupling strength of $k_S\gg 1$ is presumably most relevant for a
comparison with real pedestrian behaviour. Therefore, for many
applications the restriction to $v_{{\rm max}}=1$ is adequate to
reproduce realistic behaviour. In other cases (see above) a choice of
$v_{{\rm max}}>1$ is indispensable.

Finally, Fig.~\ref{v_fig_5}b shows the different evacuation times
for $k_D=0.0$ and varying $v_{{\rm max}}$. Since $k_S$ is very small
($k_S=0.4$), they are decreasing with $v_{{\rm max}}$. The
characteristic non-monotonous behaviour for increasing $k_D$ (see
\cite{ourpaper2}) is reproduced for all $v_{{\rm max}}$ values. Since
in this parameter regime the particle density is a restricting factor
for the evacuation time due to the clogging occuring at the exit, for
$v_{{\rm max}}\geq 2$ one finds some kind of convergence of the time
values.


\section{Finer discretisation of space}
\label{disc}

Increasing the maximal velocity of pedestrians is not the only
modification that influences the realism of the model.  One of the
boldest assumptions in CA models for pedestrian motion is the 
representation of space as a grid of cells and the corresponding restriction
of the spatial resolution to the cell size, in this case 40~cm.
However, it can be argued that this simplification is justified by the
reaction time (or decision time) which can be easily identified in
the model. Nevertheless, an -- preferably qualitative
-- investigation of the discretization effects would shed some light
on the issue.  The space discretisation used so far has been chosen in the
simplest possible way such that a pedestrian occupies only one cell.
This leads to a length scale of $a=40$ cm (see Sec.~\ref{sec:model}).
However, for some applications this might not be adequate and a finer
discretisation is necessary. In the following we will discuss the
necessity of a smaller length scale, ways of its implementation and
the consequences for the behaviour of the model.

\subsection{Motivation and consequences}

The identification of the size of one lattice site with the typical
space occupied by a pedestrian in a dense crowd is most natural
for the construction of a CA model for pedestrian dynamics. For 
square cells this leads to a cell size of $a^2=1/\rho^{\rm (emp)}_{\rm
  max}$ and $a=40$ cm where $\rho^{\rm (emp)}_{\rm max}\approx 6.25$
persons$/\text{m}^2$ is the maximal empirically observed density \cite{weid}. 
All units of space are then a multiple of the length scale $a=40$ cm. 
In contrast to the introduction of higher velocities discussed in
the previous section which was motivated by the unrealistic shape
of the fundamental diagram, there are no simulation results that
indicate the necessity of a better spatial resolution.
Nevertheless there are good reasons to introduce a finer
discretisation of space.
\begin{enumerate}
\item If the model is used not only for basic investigations of
  pedestrian dynamics, but also for the assessment of evacuation
  scenarios in complex structures (like passenger vessels or football
  stadiums), an accurate representation of the geometries is
  desirable. A finer discretisation corresponds to a more accurate
  representation of geometrical structures in a natural way. This can
  be seen from Fig.~\ref{d_fig_2}.
\begin{figure}[h]
\begin{center}
\includegraphics[width=.55\textwidth]{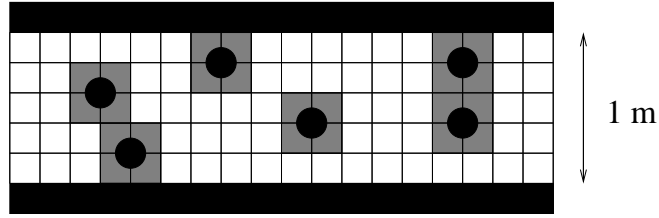}
\end{center}
\caption[]{
  Representation of a corridor of width $1$~m
  in the $20$-cm-model. Such a corridor can not be
  reproduced exactly in the $40$-cm-model. The four cells occupied by
  each particle are shaded gray.}
\label{d_fig_2}
\end{figure} 
\item As we have seen in Sec.~\ref{vervmax} some situations require the
use of higher velocities $v_{\rm max}>1$. If one wants to keep the time scale
unchanged this has to be compensated by introducing a smaller length
scale. 
\item In principle it would be interesting to consider the continuum limit
$a\to 0$ in order to make contact with models that are based on a 
continuous representation of space, e.g.\ the social force 
\cite{social,social0} and similar models \cite{TM95} or hydrodynamic 
approaches \cite{hydro2,handbook,hhydro,hughes1,hughes2}. 
\end{enumerate}

\begin{figure}[h]
\begin{center}
\includegraphics[width=.35\textwidth]{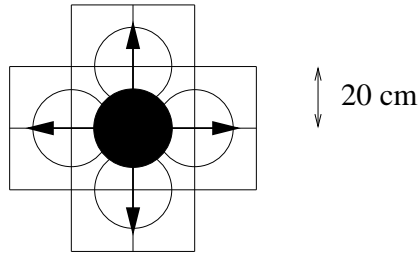}
\end{center}
\caption[]{Occupation of  $2\times 2$ lattice sites of cell length 
$a=20$ cm and possible transitions in a time step.}
\label{d_fig_1}
\end{figure}

The straightforward approach to investigate the discretization effects
is to vary the cell size.  If one chooses a cell size of $a=20$~cm, the
area $A=1/\rho_{{\rm max}}^{{\rm (emp)}}$ occupied by a pedestrian is 4 
cells (see Fig.~\ref{d_fig_2}) in order to reproduce the maximal density
$\rho_{{\rm max}}^{{\rm (emp)}}$.
The natural generalisation for the allowed
motion of the original model as described in Sec.~\ref{sec:model} 
(in the following denoted as 40-cm-model) is the
motion of the center of a pedestrian by one cell (i.e.\ a distance $a$,
see Fig.~\ref{d_fig_1}). 
This is only possible if the two neighbouring cells are unoccupied.
As can be seen from the relation (\ref{speed_max}) (with $v_{{\rm max}}=1$)
for the three fundamental quantities
the time step $\Delta t$ in the 20-cm-model is half as long as in 
the 40-cm-model\footnote {An extension of the 20-cm-model to a model with 
walking speeds $v_{\rm max}>1$ is possible in the same way as discussed in
  the previous section, but will not be considered in the following.}.
For the empirical velocity of $v_{\rm emp}=1.3$~m/s and $a=20$~cm
the time scale is $\Delta t=0.15$~sec. For an arbitrary spatial
discretisation $a$ one obtains the time scale 
$\Delta t=a/v_{{\rm max}}^{{\rm (emp)}}$.

Finer discretisations of the underlying geometric structure for
fixed velocity $v_{{\rm max}}$ therefore lead to smaller length and 
time scales of the model. So the limit $a\to 0$ results in a space {\em and}
time continuous model. This limit allows a comparison between
discrete and continuous models, but is not considered here.

The occupation of four lattice sites by one particle has further consequences
in the $20$-cm-model, especially for the floor field model described
in Sec.~\ref{sec:model}.
\begin{enumerate}
\item Every particle leaves two of the four cells it occupied
  previously (Fig.~\ref{d_fig_1}). In both these cells the dynamic 
  floor field has to be increased.
\item Since a particle occupies always four lattice sites, the
  transition probabilities are calculated by averaging the specific
  field values of static and dynamic floor fields $S$ and $D$.
\end{enumerate}

\begin{figure}[h]
a)\hspace{0.39\textwidth} b)\vspace{-0.1cm}
\begin{center}
\includegraphics[width=.23\textwidth]{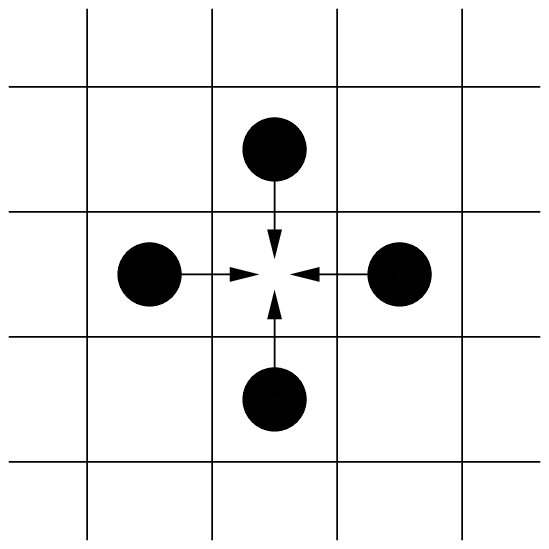}\qquad\qquad\qquad
\includegraphics[width=.45\textwidth]{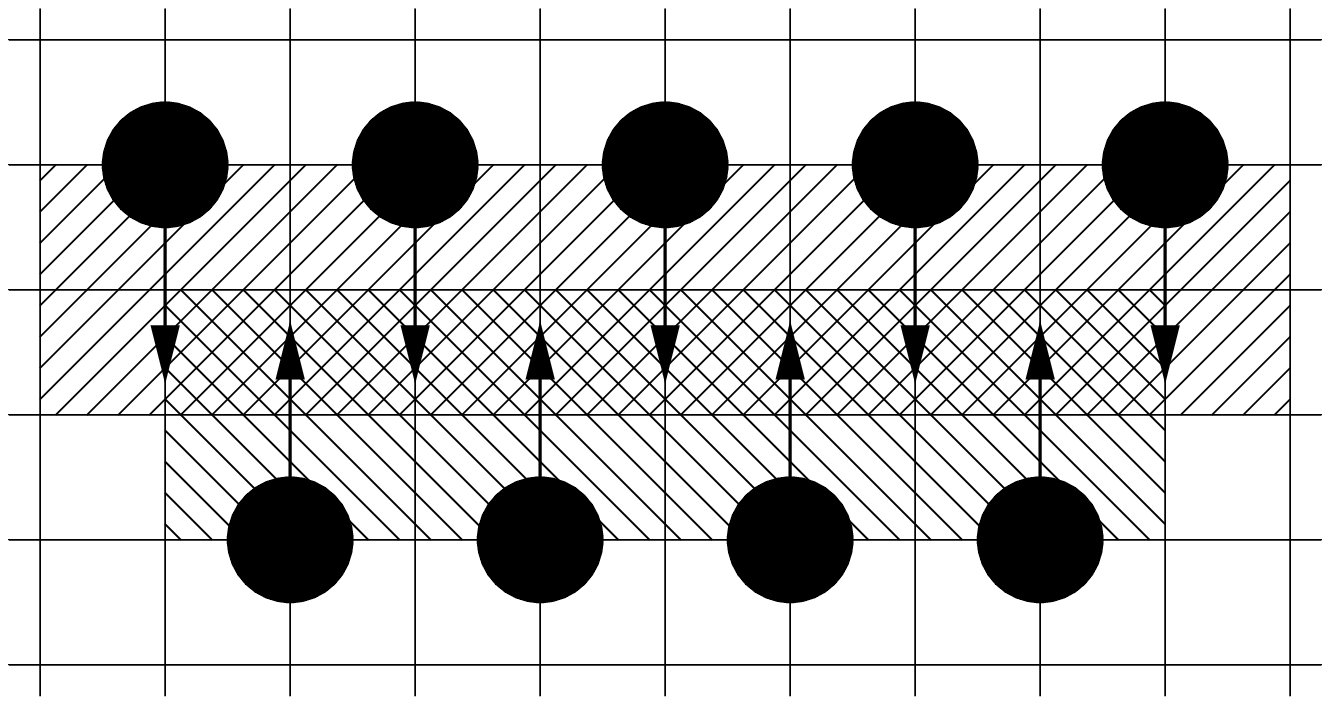}
\end{center}
\caption[]{{\bf a)} Typical local conflict in the $40$-cm-model.
{\bf b)} In the $20$-cm-model in principle all particles of a system 
can be part of a conflict. Hatched cells contribute to the non-local conflict.}
\label{d_fig_4}
\end{figure}

In any CA, not only the floor field model, an
important new effect related to conflicts occurs when the cell size is
reduced and particles occupy more than just one cell. 
In the $40$-cm-model, conflicts can be described as {\em local}
interactions (Fig.~\ref{d_fig_4}a), because only a small number of particles 
can access the
same lattice site which is determined by the coordination number
of the lattice. On the other hand, in the $20$-cm-model
conflicts are not necessarily local anymore. 
Here, conflicts are not restricted to single cells and can spread over 
a wider area of the lattice. This is illustrated for a typical
situation in Fig.~\ref{d_fig_4}b. 
Assume that the permutation $\sigma$ (see Sec.~\ref{sub_hop}) is such 
that the particles move from ``left to right", i.e. all the particles 
in the top row are allowed to move and the particles in the bottom row 
are not. Switching $\sigma^{-1}(1)$ and 
$\sigma^{-1}(2)$ will change the movement of 
all particles. This is not the case for particles that occupy only
one cell, i.e.\ $a=40$~cm.

In the floor field model conflicts allow for the introduction of
friction parameter $\mu$ \cite{KNS,KKNSS} (see Sec.~\ref{sec:model}).
Due to the possible non-local nature of conflicts in the $20$-cm-model
this concept can not straightforwardly be generalized.
New concepts for the description of conflicts are necessary, which is 
currently under investigation and will be discussed in a future publication.

In the next section simulation results of the $20$-cm-model will be
compared to results of the conventional model.

\subsection{Simulation results}

\subsubsection{Evacuation times}
\begin{figure}[h]
a)\hspace{0.485\textwidth} b)\vspace{-0.7cm}
\begin{center}
\includegraphics[width=0.495\textwidth]{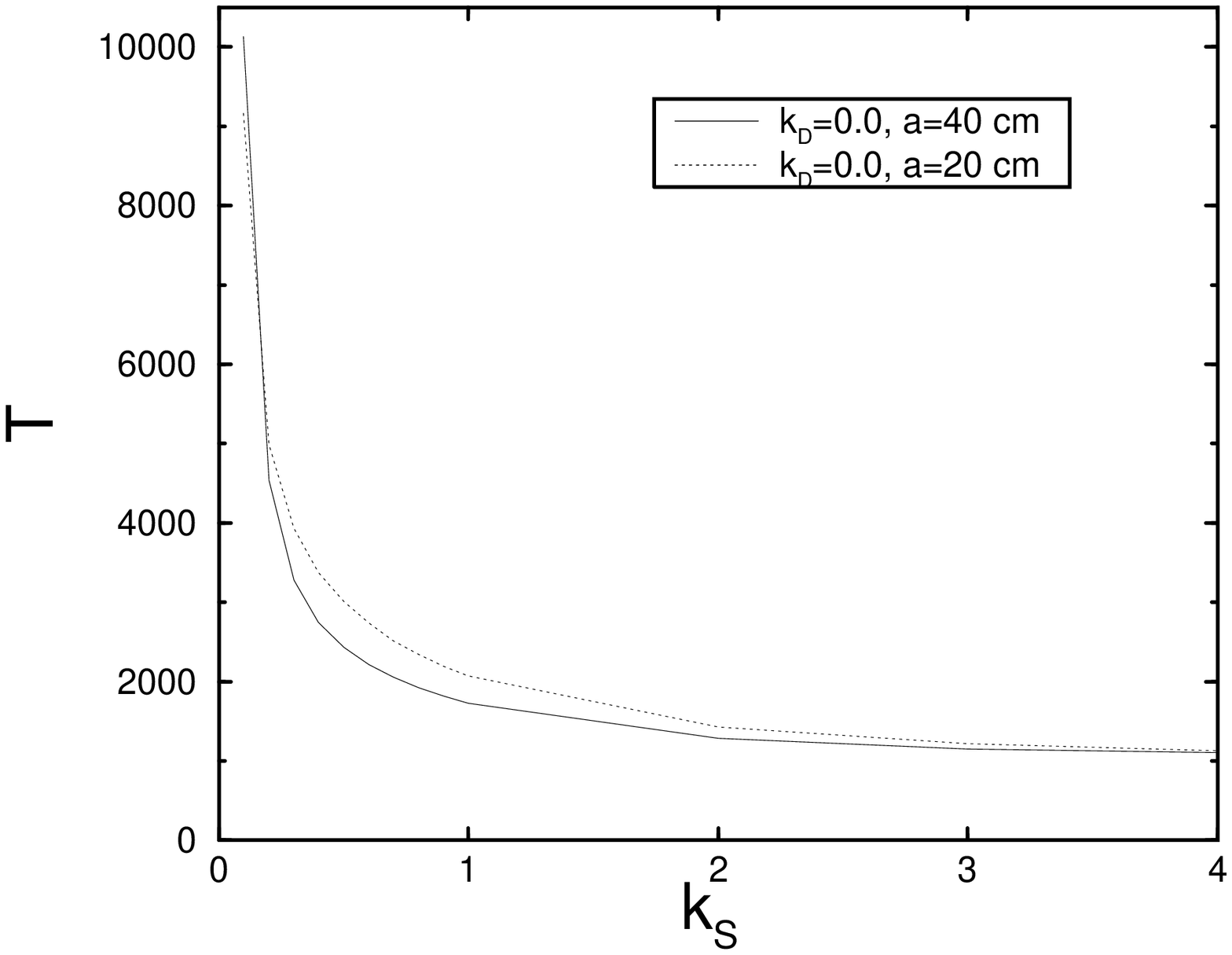}
\hfill
\includegraphics[width=0.495\textwidth]{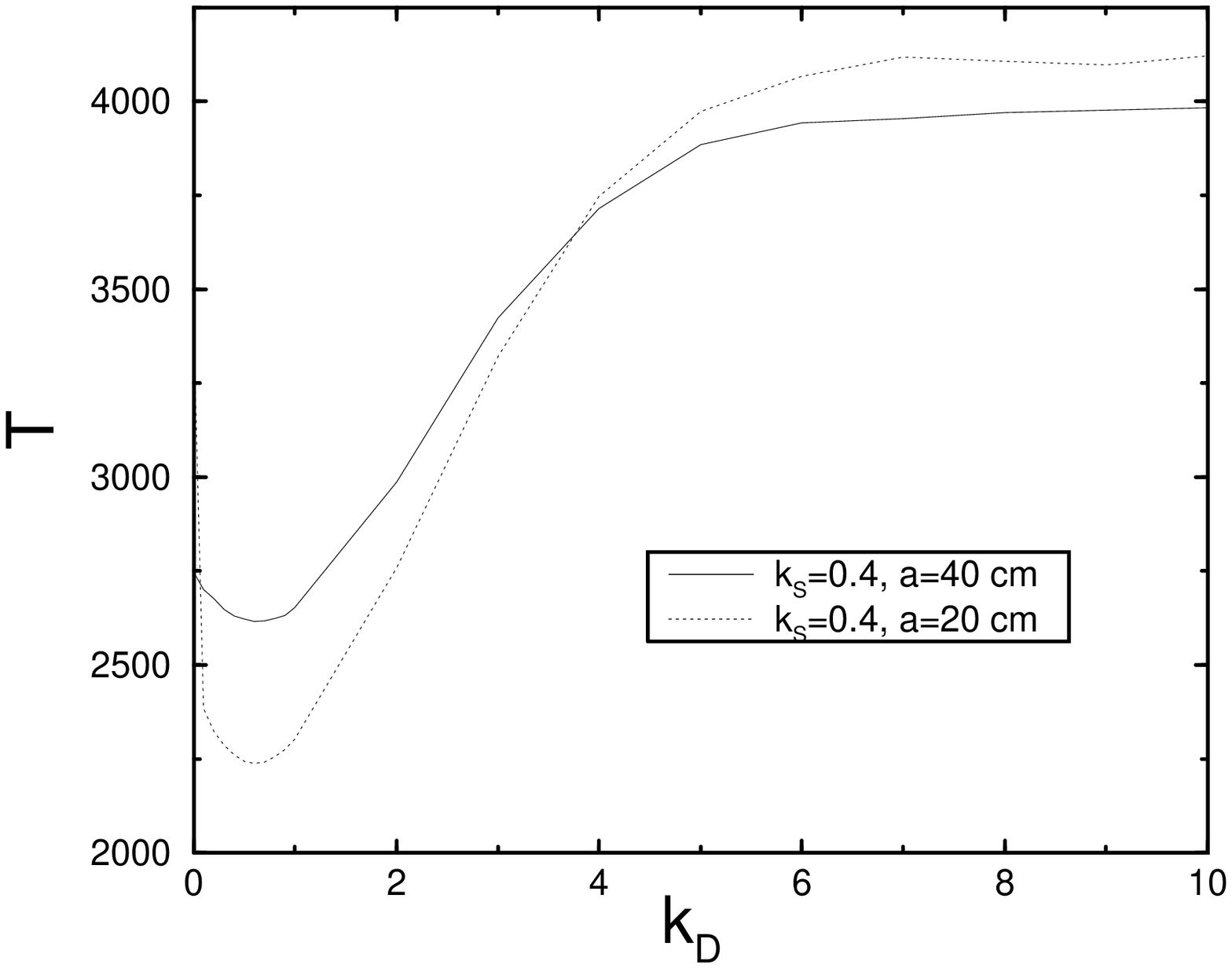}
\end{center} 
\caption{Averaged evacuation times for a room of size $62\cdot40\,{\rm cm}
\times 62\cdot40\,{\rm cm}$ and an exit of width  $80$ cm: 
{\bf a)} $k_D=0.0$ fixed and variation of $k_S$; 
{\bf b)} $k_S=0.4$ fixed and variation of  
$k_D$ ($\delta=0.3$ and  $\alpha=0.0$).}
\label{d_fig_5}
\end{figure}

First, we will turn to the influence of the discretisation on
evacuation times. We measured the averaged times of the evacuation of
a large single room with an area of about (24.8~m)$^2$, corresponding 
to $62\times 62$ cells for the $a=40$-cm-model or $124\times 124$ cells
for the $a=20$-cm-model. The particles leave the room via a door of width 
$80$ cm (two or four lattice sites, respectively). Fig.~\ref{d_fig_5}
shows the impact of the variation of the parameters $k_S$ and $k_D$ in
the $a=20$-cm and $a=40$-cm-model\footnote{The times are measured in
  update steps. Therefore, for a comparison of the two models, the
  simulation times of the $a=20$-cm-model have to be divided by two.}.
The initial particle density is $\rho=0.3$ corresponding to 1153 particles.
The qualitative behaviour of the model when varying the parameters
(see Sec.~\ref{vervmax} and \cite{ourpaper2}) is not influenced by
the finer discretisation. Because of slightly different transition
probabilities due to the averaging of the field values of four sites
for each walking direction in the $a=20$-cm-model, small deviations
only exist for coupling strengths $k_S<3$ (see Fig.~\ref{d_fig_5}a).
This also leads to different time values for the parameter
choice $k_D=0.0$ and $k_S=0.4$ in Fig.~\ref{d_fig_5}b). It is
remarkable that the effect of the non-monotonic parameter dependence
\cite{ourpaper2} of the curve is most pronounced in the $a=20$-cm-model.

\begin{figure}[h]
\begin{center}
\includegraphics[width=.4\textwidth]{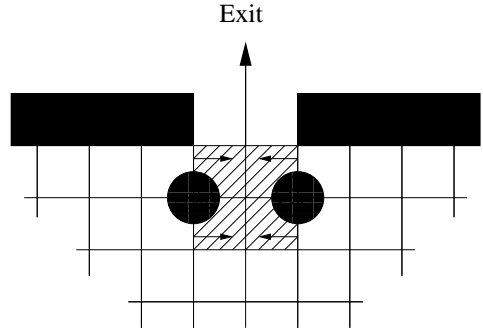}
\end{center}
\caption[]{Blockage of bottlenecks or exits of width $40$ cm in the 
$20$-cm-model caused by two particles. Both particles try to access the 
four lattice sites in front of the exit.}
\label{d_fig_7}
\end{figure}

The most important difference between the 20~cm- and 40~cm-model
is the possibility of a deadlock in front of a small exit.
Consider the flow through a bottleneck of the dimension
of the particle width (i.e.\ bottleneck and particle have the width of
two cells or $40$ cm). 
If two particles reach the bottleneck at the
same time step, their mutual blockage (see Fig.~\ref{d_fig_7})
may lead to vanishing flows. Similar observations have been
reported in experimental investigations (see e.g.\ \cite{Muir96,hbw}).
This phenomenon can occur in pushing pedestrian crowds at bottleneck-shaped 
exits of small width. It has also been observed for some continuum
pedestrian models with extended particles \cite{panic} and similarly in
granular flow through a funnel \cite{funnel1,funnel2}.
Therefore it is remarkable that it does {\em not} appear in the original
$40$-cm-model.

On the other hand, the movement itself is more 
effective for the 20~cm-model compared to the 40~cm-model.
For very high values of $k_S$, the overall evacuation time is 
determined by the outflow through the door. In the case considered here 
($w=80$~cm~$=4$ cells if $a=20$~cm), the blockage does basically not occur.
For very low values of $k_S$ (cf.~Fig.~\ref{d_fig_5}), the particles perform a 
random walk and therefore the discretization ($a=20$~cm vs.\ $a=40$~cm) does 
not play a prominent role.
This interpretation is also supported by the results shown in 
Fig.~\ref{d_fig_8} where the smallest possible exit is investigated.
In this case, the flow increases with the density for small coupling 
strengths $k_S$ and increases with large coupling strengths.
However, this does not really explain the subtle differences in the area 
$0.5\leq k_S \leq 2$, which will be investigated in more detail at a 
later point.

\subsubsection{Flow measurements}

\begin{figure}[h]
\begin{center}
\includegraphics[width=0.55\textwidth]{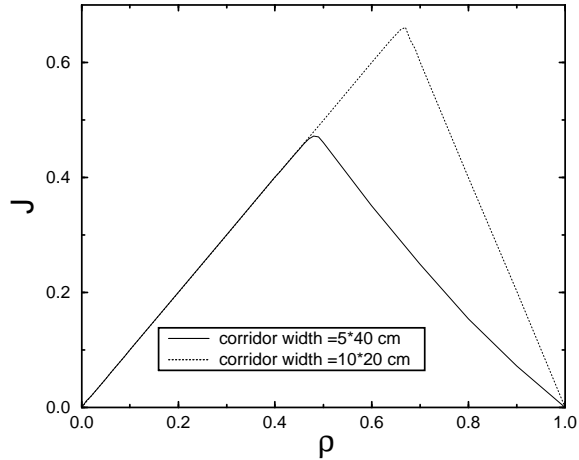}
\end{center}
\caption{Fundamental diagrams in the ordered regime ($k_S=10.0$, $k_D=0.0$) 
of both model variants. Width of corridor: $5$ cells 
in the $40$-cm-model and $10$ cells in the  $20$-cm-model.}
\label{d_fig_6}
\end{figure}

The discretisation has a significant influence on the flow. 
Fig.~\ref{d_fig_6} shows the fundamental diagrams in the ordered regime for
a corridor of width 2~m (i.e.\ width five lattice sites in the
$a=40$-cm-model and ten lattice sites in the $a=20$-cm-model).
The flows of both models are identical up to $\rho\approx\frac{1}{2}$.
The density region $\rho<\frac{1}{2}$ characterises in both models the
free flow regime. Nearly all particles can move to their right next
neighbour site in each time step, the flow increases linearly with the
density. For a density of $\rho\approx\frac{1}{2}$, the probability of
occupied next neighbour sites is highly increasing in the
$40$-cm-model, so the average velocity of the particles is decreasing
to a value $\langle v_x \rangle<1$. Therefore, the flow decreases with
increasing density and is additionally decreased by lane changing
events (see Fig.~\ref{v_fig_4}).

The behaviour is different in the $20$-cm-model: The space a particle
needs for unimpeded movement in one time step corresponds to only half
the length of its own size. Therefore, the average particle velocity
is $\langle v_x \rangle\approx 1$ up to a density of
$\rho\approx\frac{2}{3}$, which leads to a linear increase of the flow
up to this density. For $\rho>\frac{2}{3}$ one finds a sharp decrease
of the flow.

Thus the finer discretisation results in a shift of the maximum flow
to regions of higher density. This effect will be present for even
finer discretisations. As in Sec.~\ref{vervmax}, experimental data
point at a realistic flow maximum for densities $\rho<\frac{1}{2}$.
Therefore the $20$-cm-model (with $v_{\text{max}}=1$) can not reproduce 
these results qualitatively. 

\begin{figure}[h]
\begin{center}
\includegraphics[width=0.55\textwidth]{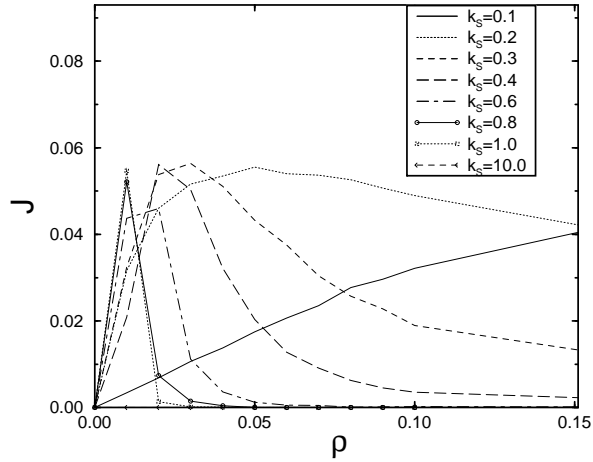}
\end{center}
\caption{Fundamental diagram of a corridor ($93\times 33$
cells of size $40~{\rm cm}\times 40~{\rm cm}$) with
a bottleneck of width $40$ cm.
The couplings to the static and dynamic floor fields are $k_S$ and $k_D=0.0$.}
\label{d_fig_8}
\end{figure}

Fig.~\ref{d_fig_8} shows the flow $J$ through a bottleneck of width
$40$ cm versus the parameters $k_S$ and $\rho$.
Only for very small $k_S$ ($k_S=0.1$) a small particle flow is
sustained for higher densities. For increasing $k_S$ the flow breaks
down at very small densities (for $k_S=10.0$ the flow is zero for all
densities). The system freezes for growing coupling strength $k_S$
to the static floor field. This is somewhat similar to the paradoxical
{\it faster-is-slower-effect} \cite{dhrev,panic}, as an increasing $k_S$
implies higher individual walking speeds. The clogging effect observed 
in our model is due to the local blockage shown in Fig.~\ref{d_fig_7}.


\section{Conclusion}

We have systematically investigated the influence of the interaction
range  $v_{{\rm max}}$ and the spatial discretisation $a$ on
the behaviour of two-dimensional CA models for pedestrian motion.
Using simulations of a corridor and evacuation from a large room
we have shown that both parameters can have a significant influence
on the properties of the model. Although we have used a specific model,
the floor field CA, for our simulations due to its ability to reproduce
the observed {\em collective} phenomena correctly we believe that
the effects discussed here are generic for any discrete approach
to pedestrian dynamics.
Of course this requires further investigations in the future.

First we have considered the extension of the interaction range which
we have implemented through a larger walking speed $v_{{\rm max}}$.
We have seen that there are different possibilities to achieve this.
This modification leads to a considerable improvement of the
agreement of fundamental diagrams for the motion along a corridor
with empirical findings. In the original case $v_{{\rm max}}=1$
the fundamental diagram of the model was almost symmetric with
respect to the density $\rho=0.5$. The reason is that for a such
a uni-directional motion where swaying is strongly surpressed many
models effectively reduce to a number of independent ASEPs\footnote{
One could expect that in the case of a mixture of species with different
$v_{{\rm max}}$ the decoupling of the lanes will be less pronounced.}
and hence the observed flow-density relation reflects its particle-hole 
symmetry. As in models of highway traffic \cite{css} the maximum in
the homogeneous case is shifted
to lower densities for larger $v_{{\rm max}}=2$ and thus leads to
a more realistic approach.

As a second scenario we have studied the evacuation from a large
room. Here it turned out that in most cases the results are not
affected strongly by the introduction of a larger velocity $v_{{\rm max}}$.

Summarizing, these results indicate that the choice $v_{{\rm max}}=1$ 
might not in all cases be justified. As indicated above the fact that for
the corridor one has to deal with a basically one-dimensional motion
is important. Here it is remarkable that also for a continuum model
like the social force model this problem requires a rather different choice 
of interaction parameters \cite{hbw} compared e.g.\ to evacuation problems,
especially concerning the repulsive interaction forces. This shows
that both model classes have to be generalized in order to achieve a proper
description of all possible situations in a unified way.

The case of the second non-trivial fundamental parameter, namely the 
cell size, is even more intricate.  A finer spatial discretisation
does not only provide a connection to a different class of microscopic 
models based on continuous representation of space. In many applications
to real life problems the geometries under consideration can not be
described by lengths that are multiples of the length scale $a=40$~cm
of the original model. We have argued that a reduction of this length
scale, and thus the cell size, must be accompanied by an increase of
the size of the particles. In order to reproduce the empirically
observed maximal density they have to occupy now more than one cell,
e.g.\ four in the case of $a=20$~cm. If the particles should have
the same symmetry as the underlying lattice this also restricts
the allowed length scales. E.g.\ for the square lattice the
most natural extensions of the particles are $2^n\times 2^n$ cells
corresponding to a length scale $a_n=\frac{40}{n}$~cm.

By studying the same two scenarios as for the influence of $v_{{\rm max}}$
we found two main consequences of using a finer discretisation. 
On the one hand, the flow is increased since the particles can now move 
ahead if there is space available less then their own size.  On the
other hand, particles can now block each other more easily.
This might lead to non-local conflicts where groups of pedestrians
mutually block each others motion (Fig.~\ref{d_fig_4}). 
Furthermore, at exits blocking 
can result in a complete breakdown of the outflow. This is not related
to the discreteness of the model and can also be observed in continuum
models with particles of finite size.

It would be interesting to perform the continuum limit of
our model to relate it to other continuum approaches. A naive continuum
version of the floor field model is currently under investigation
and might serve as a reference model. Our present study has shown
that finer discretisation or larger velocities in combination with a
discrete time (parallel) dynamics leads to the possibility
of complicated conflict situations. One could try to solve this
problem by using a continuous time dynamics, but this might
drastically increase the computational complexity of the model
in a different way.

Even though our investigations are based on a square grid (one could
also think of a hexagonal lattice, and indeed there are models based
on this \cite{chopard,maniccam}) the results obtained are generally valid. 
As a next step currently the
combined effect of larger $v_{{\rm max}}$ and smaller cells is
investigated. Then one can expect to encounter a combination of many
effects observed here. However, e.g.\ in the case of flow along a
corridor these modifications have opposite quantitative effects.
Whereas an increase of $v_{{\rm max}}$ tends to shift the location of
maximal flow to smaller densities the reduction of the length scale
has the opposite effect. This work is in progress and will be reported
elsewhere.

Summarizing we can say that discrete models like cellular automata
yield quite realistic results. E.g.\ the floor field model is able to 
reproduce collective effects (such as lane formation in counterflow
etc.) observed empirically. However, for applications where also a reliable
{\em quantitative} prediction should be made one has to take some
care in the choice of the basic parameters as cell size and
interaction range.


\end{document}